# Exactly solvable model for transmission line with artificial dispersion

A.B. Shvartsburg,[1,2] S.N. Artekha,[2,a)] N.S. Artekha[3]

AFFILIATIONS

[1] *Joint Institute for High Temperatures of RAS, Moscow 125412, Russia*

[2] *Space Research Institute of RAS, Moscow 117997, Russia*

[3] *Peoples' Friendship University of Russia, Moscow 117198, Russia*

[a)] Author to whom correspondence should be addressed: sergey.arteha@gmail.com

ABSTRACT

The problem of the emergence of wave dispersion due to the heterogeneity of a transmission line (TL) is considered. An exactly solvable model helps to better understand the physical process of a signal passing through a non-uniform section of the line and to compare the exact solution and solutions obtained using various approximate methods. Based on the transition to new variables, the developed approach made it possible to construct exact analytical solutions of telegraph equations with a continuous distribution of parameters, which depend on the coordinate. The flexibility of the discussed model is due to the presence of a number of free parameters, including two geometric factors characterizing the lengths of inhomogeneities in values of the inductance *L* and of the capacitance *C*. In the new variables, the spatiotemporal structure of the solutions is described using sine waves and elementary functions, and the dispersion is determined by the formulas of the waveguide type. The dispersive waveguide-like structure characterized by refractive index *N* and cut-off frequency *Ω*. The exact expressions for the complex reflection and transmission coefficients are derived. These expressions describe phase shifts for reflected and transmitted waves. The following interesting cases are analyzed: the passage of waves without phase change, the reflectionless passage of waves, and the passage of signals through a sequence of non-uniform sections. The developed mathematical formalism can be useful for the analysis of a wider range of problems.

## I. INTRODUCTION

In recent years, studies of the interaction of waves with inhomogeneous media have been actively carried out (see[1-8] and references therein). The notion of a "barrier" is frequently used in various fields of science, for which a wave description can be applied, for example, in plasma physics (wave barriers), in solid-state physics (quantum-mechanical barriers), in optics, etc. A segment of a non-uniform transmission line can also represent a barrier with respect to propagating waves. Exactly solvable models are interesting for science, technology and education, since they can help in purposeful scientific search and in understanding of phenomena under investigation. A key role in many phenomena is played by a variety of resonant effects. The study of such resonant effects is of considerable interest for various practical applications. Exactly solvable models can help in their search, for example, when analyzing the possibilities of realization for reflectionless passage of waves through wave barriers. Such a model will be presented in the article.

We consider the classical "telegraph equations"[9-12] describing the voltage $U$ and current $I$ in the transmission line (TL). For example, for a homogeneous stripline formed by two metal strips and a dielectric layer between them, the inductance and capacitance per unit length will be:[9] $L = 4\pi \cdot 10^{-7} a/b$ [H/m], $C = 10^{-9} \varepsilon_r b/(36\pi a)$ [F/m], where $\varepsilon_r$ is an effective dielectric constant of the layer, $a$ – its thickness, $b$ – its width. However, we will consider the non-uniform TL with continuously distributed capacity $C(z)$ and inductance $L(z)$ per unit length. Let us model these distributions of capacity and inductance along the line by means of products $C(z) = C_0 P^2(z)$ and $L(z) = L_0 W^2(z)$, where $P(z)$ and $W(z)$ are some dimensionless real smooth functions. The telegraph equations for the lossless TL with these parameters can be written (in SI system) as

$$\frac{\partial U}{\partial z} + L_0 W^2(z) \frac{\partial I}{\partial t} = 0, \quad \frac{\partial I}{\partial z} + C_0 P^2(z) \frac{\partial U}{\partial t} = 0. \tag{1}$$

It is known that the analysis of group theory can be attributed to the most general methods for the analysis of partial differential equations.[13-17] Effective methods for integrating equations of mathematical physics include methods based on knowledge of their continuous, point, nonlocal, or potential symmetries. These methods are successfully used to analyze general telegraph and wave equations with variable coefficients.[18-21] It is known that the number of dependencies for the coefficients that admit an analytical solution of such equations is limited.[2,9,22] In our analysis, we will follow the ideas developed in the works.[2,23-28] For this, introducing the generating function $\Psi$:

$$I = -\frac{\partial \Psi}{\partial t}, \quad U = \frac{1}{C_0 P^2(z)} \frac{\partial \Psi}{\partial z}, \tag{2}$$

we can reduce the system (1) to one equation

$$\frac{\partial^2 \Psi}{\partial z^2} - \frac{W^2(z) P^2(z)}{v_0^2} \frac{\partial^2 \Psi}{\partial t^2} = \frac{2}{P(z)} \frac{\partial P}{\partial z} \frac{\partial \Psi}{\partial z}, \quad v_0^2 = \frac{1}{L_0 C_0}. \tag{3}$$

Next we will use the method of phase coordinates:[2,23] using of new variable

$$\theta = \int_0^z P^2(z') dz', \tag{4}$$

permits one to eliminate the right side from Eq. (3)

$$\frac{\partial^2 \Psi}{\partial \theta^2} - \frac{1}{v_0^2} \frac{W^2(z)}{P^2(z)} \frac{\partial^2 \Psi}{\partial t^2} = 0. \tag{5}$$

As it can be seen from equation (5), there is a certain analogy between the microwave phenomena studied here and optical phenomena in inhomogeneous media with spatial modulation, as well as with the phenomenon of Alfvén wave propagation through inhomogeneous plasma.[2,4] Consequently, the mathematical formalism that will be developed in the work, and the results obtained may be useful for the analysis of a wider range of problems.

The main purposes of this article are to describe the appearance of dispersion in a non-uniform transmission line using a precisely solvable model and to derive the rigorous expressions for the complex reflection and transmission coefficients. These expressions describe a change in the amplitudes of transmitted and reflected waves, as well as the formation of their phase shifts. This will allow us to analyze important particular cases of signal propagation, as well as highlight possible new properties of long transmission lines that may have practical application. The analysis will also touch on the possibility of constructing a transmission line in which several non-uniform segments are combined.

The article is organized as follows: Sec. II discusses a precisely solvable model for a non-uniform TL with distributed parameters; Sec. III studies the propagation and tunneling of current and voltage waves, and analyzes a number of important special cases; finally, Sec. IV contains conclusions.

## II. DISPERSION OF NON-UNIFORM TRANSMISSION LINE WITH CONTINUOUSLY DISTRIBUTED PARAMETERS (EXACTLY SOLVABLE MODEL)

Till now the distributions $W(z)$ and $P(z)$ were arbitrary regular functions. The number of dependencies for the coefficients that allow an analytical solution of (5) is finite. It is needed to choose such a dependence in (4) that allows the explicit expression of the variable $z$ through the variable $\theta$. We consider the following model of TL with distributed capacity $P(z)$ and inductance $W(z)$:

$$P(z) = \left(1 + \frac{z}{l_1}\right)^{-1}, \quad W(z) = \left(1 + \frac{z}{l_2}\right)^{-1}. \tag{6}$$

The characteristic lengths $l_1$ and $l_2$ in (6) are the free parameters of the discussed model of non-uniform TL; positive and negative values of these quantities correspond to growth or decrease of capacity and inductance along the TL. This choice provides a fairly wide range of possibilities for modeling various heterogeneities. As it can be seen from Fig. 1, the change in the parameters on the non-uniform section of the transmission line can occur with respect to the homogeneous section both continuously and abruptly, and herewith the inductance or capacitance can independently either increase, or decrease, or remain constant.

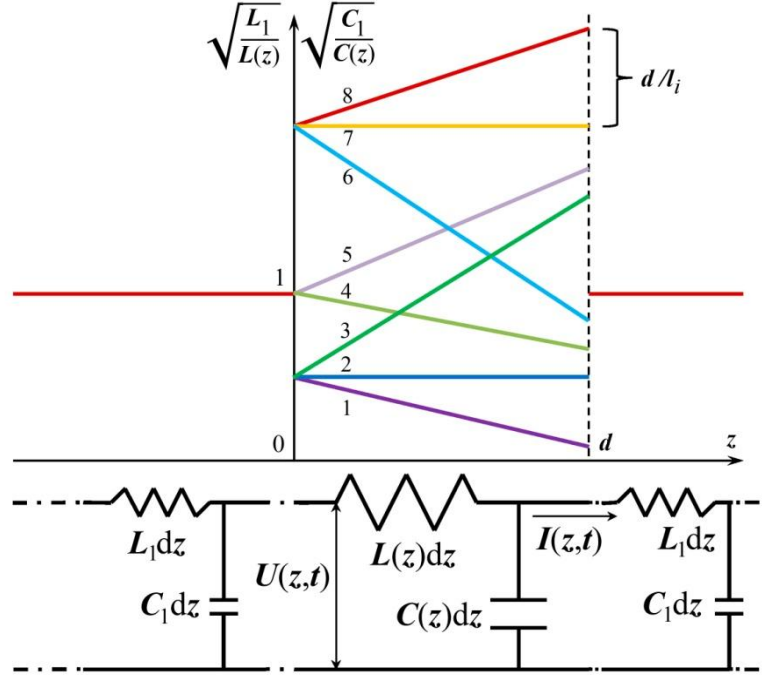

**Fig. 1.** The equivalent circuit of different sections d$z$ for the transmission line with a non-uniform section [0, $d$] is shown in the lower figure; and some possible dependencies of dimensionless ratio of parameters are shown in the upper graphic. The definition $d/l_i$ can be seen for the straight line 8. The change in the parameters for a non-uniform section of the transmission line can occur with respect to the homogeneous section both continuously (straight lines 4 and 5) and abruptly (straight lines 1, 2, 3, 6, 7, 8), and herewith either increase (straight lines 3, 5, 8), or decrease (straight lines 1, 4, 6), or remain constant (straight lines 2 and 7).

Substitution of function $P(z)$ from (6) to Eq. (4) brings the link between the variables $z$ and $\theta$:

$$z = \theta\left(1 - \frac{\theta}{l_1}\right)^{-1}, \qquad \theta = z\left(1 + \frac{z}{l_1}\right)^{-1}. \tag{7}$$

Finally, presentation of the factor $W^2(z)P^{-2}(z)$ in Eq. (5) via the variable $\theta$ permits one to rewrite this equation in a form

$$\frac{\partial^2 \Psi}{\partial \theta^2} - \frac{1}{v_0^2}\left(1 + \frac{\theta}{l_3}\right)^{-2} \frac{\partial^2 \Psi}{\partial t^2} = 0, \qquad \frac{1}{l_3} = \frac{1}{l_2} - \frac{1}{l_1}. \tag{8}$$

Here $l_3$ is a new spatial scale; subject to the correlation between scales $l_1$ and $l_2$, the values of $l_3$ can be both positive or negative. A method for solution of equation (8) will be considered below in detail.

Inspection of Eq. (8) shows, that an unknown function $\Psi$ obeys a wave equation with a coordinate-dependent speed of wave propagation. Equation (8) can be solved by introducing a new variable $\eta$ and a new function $F$:

$$\eta = l_3 \ln\left|1 + \frac{\theta}{l_3}\right|, \qquad \Psi = F\sqrt{\left|1 + \frac{\theta}{l_3}\right|}. \tag{9}$$

Assuming, that the time dependence of function $\Psi$ is harmonic $(\Psi \propto \exp(-i\omega t))$, and using (9) we obtain from Eq. (8) the simple equation with constant coefficients governing the behavior of function $F$ in $\eta$-space:

$$\frac{\partial^2 F}{\partial \eta^2} + q^2 F = 0, \quad q = \frac{\omega}{v_0} N, \quad N^2 = 1 - \frac{\Omega^2}{\omega^2}, \quad \Omega^2 = \frac{v_0^2}{4l_3^2}. \tag{10}$$

Equation (10) can be viewed as the standard equation describing the propagation of wave with wave number $q$ in dispersive waveguide-like structure characterized by refractive index $N$ and cut-off frequency $\Omega$. The latter frequency $\Omega$ separates the frequency region where the tunneling regime is observed from the region with the wave regime of propagation. Note, that this cut-off frequency is determined by the parameters of spatial distributions of capacity and inductance (8). Equation (10) describes the propagation of a pulse with an arbitrary shape $F(t - \eta/q)$ in a space $(\eta, t)$ without distortion; as this takes place, the dependence $\eta(z)$ describes the controlled deformation of the pulse in the physical space $(z, t)$. Substitution of solutions of Eq. (10), presented via the exponents $\exp(\pm i q \eta)$, to (9) brings the exact analytical solution of Eq. (8) for the generating function $\Psi$

$$\Psi = A\sqrt{\left|1 + \frac{\theta}{l_3}\right|}\left[\exp(iq\eta) + Q\exp(-iq\eta)\right]\exp(-i\omega t). \tag{11}$$

Here $A$ and $Q$ are the constants, which have to be determined from the boundary conditions on the ends of TL. The obvious expressions for variables $\theta$ and $\eta$ through the variable $z$, obtained from (7) and (9), read as

$$\theta = z\left(1 + \frac{z}{l_1}\right)^{-1}, \qquad \eta = l_3 \ln\left|\left(1 + \frac{z}{l_2}\right)\left(1 + \frac{z}{l_1}\right)^{-1}\right|. \tag{12}$$

The exact generating function $\Psi$ (11) will be used below for computation of reflection and transmission of current and voltage waves in the non–uniform TL with artificial dispersion for the variety of physically meaningful cases stipulated by the interplay of positive and negative values of free parameters $l_1$ and $l_2$. Moreover, this approach provides the platform for comparison of eikonal and antieikonal approximations with the exact solutions.[2,29]

## III. PROPAGATION AND TUNNELING OF CURRENT AND VOLTAGE WAVES IN NON-UNIFORM TRANSMISSION LINE

Voltage $U$ and current $I$ in the TL can be found due to substitution of generating function $\Psi$ (11) to the equalities (2):

$$U = \frac{\Psi}{2C_0 l_3}\left(1+\frac{\theta}{l_3}\right)^{-1}\left[1+2iql_3\frac{\exp(iq\eta)-Q\exp(-iq\eta)}{\exp(iq\eta)+Q\exp(-iq\eta)}\right], \qquad I = i\omega\Psi. \qquad (13)$$

We consider the reflection and transmission of monochromatic wave from the segment of non-uniform TL with length $d$ described by the distributions (6) and installed into the uniform line characterized by the constant capacitance $C_1$ and inductance $L_1$ per unit length (see Fig. 1). Note, that the voltage $U_1$ and current $I_1$ describing the incident wave in the uniform area $z<0$ with capacitance $C_1$ and inductance $L_1$ can be written by means of generating function $\Psi_1$ by means of presentation (2):

$$\Psi_1 = A_1\exp\left[i(k_1 z - \omega t)\right], \quad U_1 = \frac{ik_1}{C_1}\Psi_1, \quad I_1 = i\omega\Psi_1, \quad k_1 = \omega\sqrt{L_1 C_1}. \qquad (14)$$

Taking into account the reflected wave, we can write

$$\begin{aligned}I_1 &= i\omega A_1\exp(-i\omega t)\left\{\exp[ik_1 z]+R\exp[-ik_1 z]\right\},\\ U_1 &= \frac{ik_1 A_1}{C_1}\exp(-i\omega t)\left\{\exp[ik_1 z]-R\exp[-ik_1 z]\right\}.\end{aligned} \qquad (15)$$

The problem of propagation of the current wave through a non-uniform section is analogous to the task of passage of a wave through a gradient barrier (see Fig. 2).

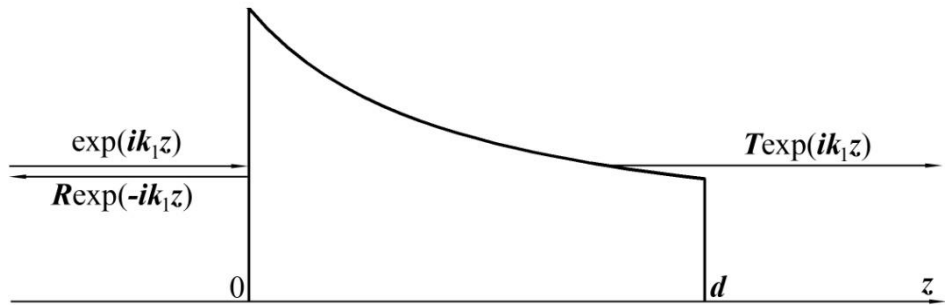

**Fig. 2.** Tunneling of a current wave (normalized to the amplitude of the incident current wave) through a gradient barrier, which is similar to an inhomogeneous section of the transmission line shown in Fig. 1. Here we see the wave incident on the [0, $d$] section, the reflected wave and the transmitted wave.

To determine the reflection coefficient $R$ of this segment for the wave incident from the area $z<0$ on the end of segment $z=0$, one has to use the continuity conditions for voltage and current at the point $z=0$. Keeping in mind the values of new variables $\theta$ and $\eta$ at the end $z=0$, $\theta(0)=\eta(0)=0$, we find from (13) – (15) the equation governing the complex reflection coefficient $R$:

$$\frac{1+R}{1-R}=\frac{Z_1}{Z_0 BN}, \quad Z_1=\sqrt{\frac{L_1}{C_1}}, \quad Z_0=\sqrt{\frac{L_0}{C_0}}, \quad B=\frac{1-Q}{1+Q}-\frac{i}{2ql_3}, \quad (16)$$

where $Z_1$ and $Z_0$ – are the impedances of uniform and non-uniform parts of TL.

For the transmitted wave, we can write

$$I_1=i\omega A_1 T\exp[i(k_1 z-\omega t)], \quad U_1=\frac{ik_1 A_1}{C_1}T\exp[i(k_1 z-\omega t)]. \quad (17)$$

The unknown parameter $Q$ in Eq. (16), determined from the continuity condition at the end of segment $z=d$, reads as

$$Q=\exp(2iq\eta_0)\frac{[2iql_3(NZ_0-\varsigma Z_1)+NZ_0]}{[2iql_3(NZ_0+\varsigma Z_1)-NZ_0]}, \quad \varsigma=\left|1+\frac{\theta(d)}{l_3}\right|=\left|\frac{1+d/l_2}{1+d/l_1}\right|, \quad \eta_0=l_3\ln\varsigma. \quad (18)$$

Substitution of $Q$ from (18) into (16) brings the value of quantity $B$

$$B=\frac{Z_0 Ns(4q^2 l_3^2+1)+2iql_3(2ql_3 Z_1\varsigma-Z_1 s\varsigma)}{2ql_3(2iql_3 Z_0 N+2ql_3 Z_1 s\varsigma+iZ_0 Ns)}, \quad s=\text{tg}(q\eta_0). \quad (19)$$

Finally, substitution of the expression of $B$ from (19) into Eq. (16) yields the complex reflection coefficient of non-uniform segment of TL:

$$R=\frac{Z_1-Z_0 BN}{Z_1+Z_0 BN}. \quad (20)$$

Quantities $k_1$ and $\eta_0$ are defined in (14) and (18) respectively. The transmission coefficient is

$$T=\sqrt{\varsigma}\exp(-ik_1 d)\{\exp(iq\eta)+Q\exp(-iq\eta)\}\frac{(1+R)}{(1+Q)}. \quad (21)$$

Note, that in the limiting case, when the distributions of capacity and inductance are coinciding $l_1=l_2$, we have $l_3\to\infty$, $\Omega\to 0$, $N\to 1$, and expression (20) is reduced to the standard formula describing the reflection coefficient of uniform segment installed to TL:

$$R=\frac{(Z_1^2-Z_0^2)\text{tg}(qd)}{(Z_1^2+Z_0^2)\text{tg}(qd)+2iZ_0 Z_1}. \quad (22)$$

Introducing the dimensionless values of $\tilde{l}_3=l_3/l_0$, $\tilde{\omega}=\omega/\Omega_0$, $\Omega_0=1/(2l_0\sqrt{L_0 C_0})$ and the notation $\tilde{Z}=Z_0/Z_1$ in (20), the ultimate results for the reflection coefficient will be:

$$R=\frac{s[\tilde{\omega}^2\tilde{l}_3^2(\varsigma-N^2\tilde{Z}^2)-\tilde{Z}^2]+i\tilde{\omega}\tilde{l}_3\tilde{Z}[s(1+\varsigma)+\tilde{\omega}\tilde{l}_3 N(1-\varsigma)]}{s[\tilde{\omega}^2\tilde{l}_3^2(\varsigma+N^2\tilde{Z}^2)+\tilde{Z}^2]+i\tilde{\omega}\tilde{l}_3\tilde{Z}[s(1-\varsigma)+\tilde{\omega}\tilde{l}_3 N(1+\varsigma)]}, \quad (23)$$

where $N = \sqrt{1 - 1/(\tilde{\omega}^2 \tilde{l}_3^2)}$, $s = \text{tg}(0.5\tilde{\omega}\tilde{l}_3 N \ln \varsigma)$, $\varsigma = \left|(1+1/\tilde{l}_2)\big/(1+1/\tilde{l}_1)\right|$, $\tilde{l}_1 = l_1/l_0$, $\tilde{l}_2 = l_2/l_0$.

For example, the value of $l_0 = d$ can be chosen for graphical presentation. We will not write out explicitly the expression for the transmission coefficient $T$ due to its bulkiness, but it can be analytically obtained on PC and analyzed in numerical computations.

We see that in the general case there are dispersion of values of $R(\omega)$ and $T(\omega)$, i.e. both values $U$ and $I$ – are the ω-dependent ones. There occurs the full reflection $R = 1$ at $N = 0$, i.e. when the wave frequency ω is equal to the cut-off frequency $\Omega$. If the propagating wave has a frequency $\omega$ less than this value of $\Omega$, the segment of non-uniform TL represents the opacity area (in this case, the wave can tunnel with damping for some short distance, but the most part of the wave will reflect). The dependence of dimensionless value of the cut-off frequency $\tilde{\Omega} = \Omega/\Omega_0$ on the dimensionless values $\tilde{l}_1 = l_1/d$ and $\tilde{l}_2 = l_2/d$ is shown in Fig. 3, where $\Omega_0 = 1/(2d\sqrt{L_0 C_0})$. Wave solutions will be observed at frequencies lying above the surface depicted here.

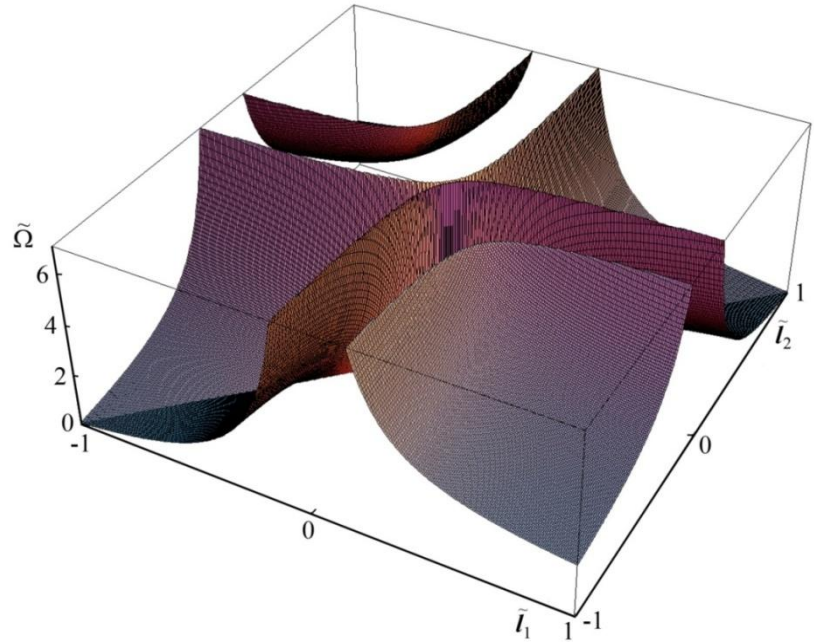

**Fig. 3.** The dependence of the dimensionless cut-off frequency $\tilde{\Omega} = 2\Omega d\sqrt{L_0 C_0}$ (see expression (10)) on the dimensionless parameters $\tilde{l}_1 = l_1/d$ and $\tilde{l}_2 = l_2/d$. An opacity region lies below the depicted surface; and the region of wave solutions exists above this surface.

Further we describe other special cases.

I. If $\text{Im}[R] = 0$ for a wave with some frequency $\omega$, then the wave will be reflected without changing its phase. The behavior of the function $\text{Im}[R]$ is characterized by the following features.

a) There are sections (bands of opacity) continuous in frequency $\omega$ and parameter $l_3$, where the imaginary part $\text{Im}[R] \approx 0$; as well as there are continuous lines connecting the values $\omega$ and $l_3$, where $\text{Im}[R] \equiv 0$, i.e. the phase of such the reflected waves remain constant.

b) With an increase in $\tilde{Z}$, the beginning of such the band slightly shifts toward somewhat higher frequencies.

c) The "frequency of oscillations" of the function $\text{Im}[R]$ with respect to the value of $\omega$ is greater for negative parameter $l_2$ than for positive parameter $l_2$.

d) An increase in the parameter $l_1$ for a fixed $\tilde{Z}$ leads to an increase in the "frequency of such oscillations" of the function $\text{Im}[R]$ depending on the values of $\omega$.

e) In the general case, for each region, a continuous change in the parameters leads alternately to an increase in the amplitude of the oscillations to unity (function $-1 < \text{Im}[R] < 1$) and to a decrease in the amplitude of the oscillations to almost zero ($\text{Im}[R] \to 0$).

For example, the dependence of the function $\text{Im}[R]$ on the dimensionless parameters $\tilde{l}_2$ and $\tilde{\omega}$ at $\tilde{Z} = 0.1$ and $\tilde{l}_1 = 0.1$ is shown in Fig. 4.

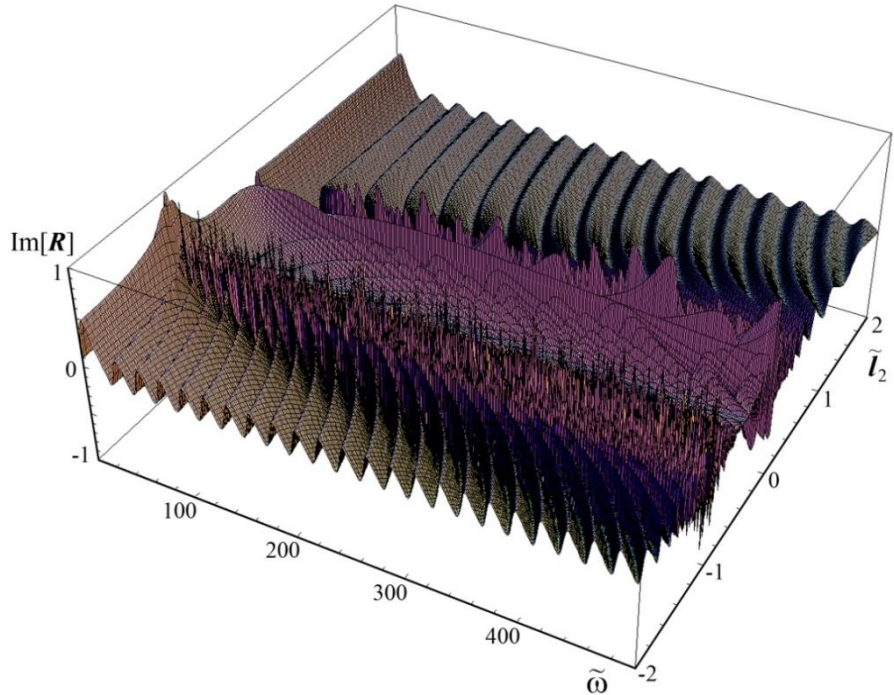

**Fig. 4.** Dependence of the imaginary part of the reflection coefficient $R$ from the expression (23) on the dimensionless parameters $\tilde{\omega}$ and $\tilde{l}_2$ at the fixed parameters $\tilde{Z} = 0.1$ and $\tilde{l}_1 = 0.1$. From this graph, it can be seen areas of continuous change in parameters, for which the equality $\text{Im}[R] = 0$ remains true (section by this plane), that is, the phase of the reflected waves does not change at all.

II. If for some wave frequency $\omega$ we have $\text{Im}[T]=0$, then the wave will be passed through the non-uniform TL without changing its phase. The behavior of the function $\text{Im}[T]$ is characterized by the following features.

a) There are sections (bands of opacity) continuous in frequency $\omega$ and parameter $l_3$, where the imaginary part $\text{Im}[T] \approx 0$; as well as there are continuous lines connecting the values $\omega$ and $l_3$, where the imaginary part $\text{Im}[T]=0$ exactly, i.e. the phase of such the forward (passed) waves remain constant.

b) With an increase in $\tilde{Z}$, the beginning of such the band slightly shifts toward somewhat higher frequencies.

c) The "frequency of oscillations" of the function $\text{Im}[T]$ with respect to the value of $\omega$ is greater to the right of the band (for negative parameter $l_2$) than to the left of the band (for positive parameter $l_2$).

d) An increase in the parameter $l_1$ for a fixed $\tilde{Z}$ leads to an increase in the "frequency of oscillations" of the function $\text{Im}[T]$ depending on the values of $\omega$.

e) In the general case, for each region, a continuous change in the parameters leads alternately to an increase in the amplitude of the oscillations of the function $\text{Im}[T]$ and to a decrease in the amplitude of the oscillations of the function $\text{Im}[T]$.

f) Since the function $\text{Im}[T]$ depends also on $C_1/C_0$, frequency of oscillations (increases with $C_1/C_0$) for this function can differ from analogous values for functions $\text{Im}[R]$ and $|T|$.

As an example, the dependence of the function $\text{Im}[T]$ on the dimensionless parameters $\tilde{l}_2$ and $\tilde{\omega}$ at $\tilde{Z}=0.1$, $\tilde{l}_1=0.1$ and $C_1/C_0=0.1$ is shown in Fig. 5.

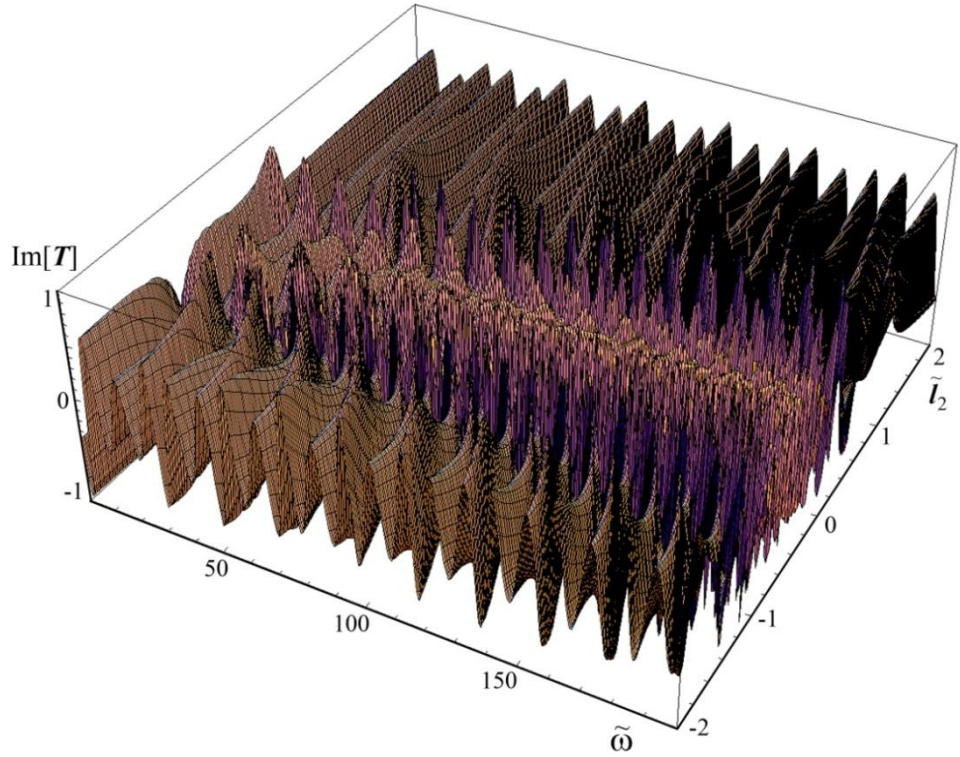

**Fig. 5.** Dependence of the imaginary part of transmission coefficient $T$ from the expression (21) on the dimensionless parameters $\tilde{\omega}$ and $\tilde{l}_2$ at the fixed parameters $\tilde{Z}=0.1$ and $\tilde{l}_1=0.1$. From this graph, it can be seen areas of continuous change in parameters, for which the equality $\text{Im}[T]=0$ remains true (section by this plane), that is, the phase of the transmitted waves does not change at all.

III. In the case when the value $T(\omega_r)=1$ (or the value $R(\omega_r)=0$, which is the same case), the phenomenon of reflectionless passage takes place for the wave with such a resonant frequency $\omega_r$. In the general case, when changing the system parameters, the amplitude of the oscillations of the function $|T|$ with frequency $\tilde{\omega}$ can alternatively increase and decrease. The phenomenon of reflectionless propagation of wave exists not for all values of the system parameters. But if there is a resonant frequency $\omega_r$ for a certain set of system parameters, then there will be infinitely many such resonant frequencies $\omega_{ri}$. As some example, the dependence of the function $|T|$ on the dimensionless parameters $\tilde{l}_2$ and $\tilde{\omega}$ at $\tilde{Z}=5.0$ and $\tilde{l}_1=10.0$ is shown in Fig. 6.

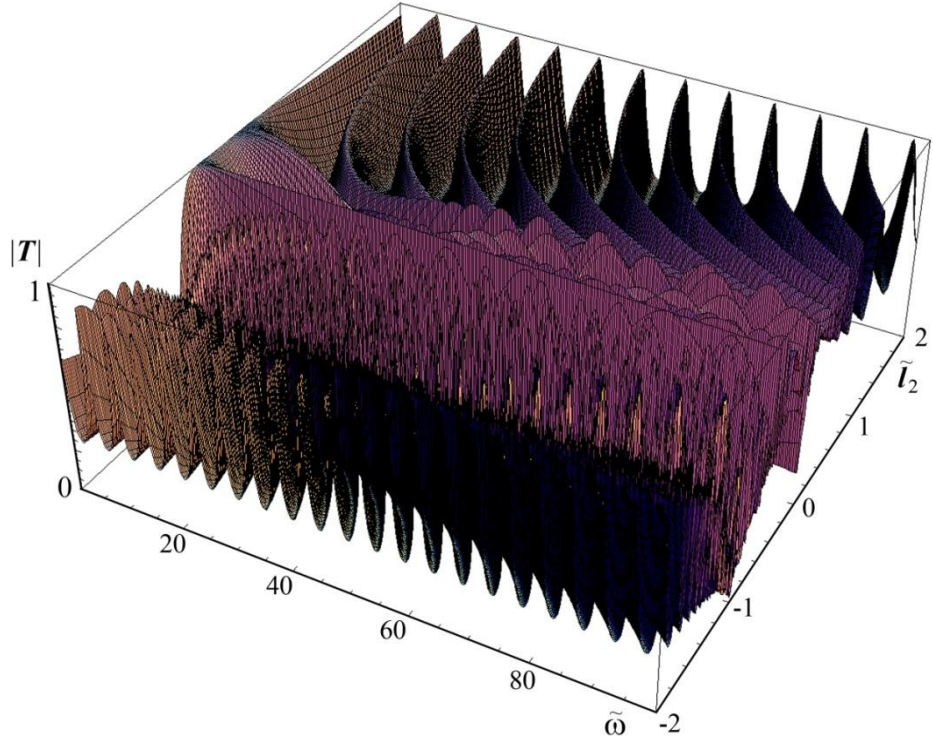

**Fig. 6.** Dependence of the absolute value of the transmission coefficient $T$ (computed module of expression (21)) on the dimensionless parameters $\tilde{\omega}$ and $\tilde{l}_2$ at the fixed parameters $\tilde{Z} = 5.0$ and $\tilde{l}_1 = 10.0$. The dependence is oscillatory in nature: there are the continuous values of parameters, for which $|T| \approx 0$, and there exist resonance frequencies, for which $|T(\omega_r)| = 1$ (resonance lines are also continuous).

If some signal is compounded of waves with resonant frequencies $\omega_{ri}$ only, then such signal ( $\sum_i \psi_i(\omega_{ri})$ ) will pass through the non-uniform TL without changing its form (no distortion). Note that to transmit information without distortion, you can use the amplitude modulation of the signals at these selected frequencies. If some signal represents the very narrow wave packet constituted of waves near the resonant frequency $\omega_r$, then such signal will pass through the non-uniform TL with a minimal changing its form. However in the general case (for an arbitrary wave packet, for example), the form of the propagating signal can change significantly (similarly for the reflected signal).

We also note that the problems considered in this article have certain analogies with some tasks that are solved in the theory of waveguides and antennas.[30-47] So, for example, reflectionless tunneling can occur in a homogeneous layer of the waveguide, which is bounded on the sides by curved boundaries. In this case, the interference of partial waves can lead to the complete suppression of the reflected radio waves. As a result, waves with wavelength exceeding the width of the slit by several times, tunnel without reflection through the narrowing of the waveguide.[30,32] Coatings (for example, from metamaterials) can be also applied for amplification of evanescent electromagnetic waves (tunneling).[32,38] For waveguides, there exists the lowest frequency – the cutoff frequency,[35] similar to the expression (10); and the wave mode less than this cannot propagate. To reduce the cutoff frequency, baffles are sometimes used.[34] Various types (modes) of waves can be excited in waveguides. Sometimes it is necessary to suppress unwanted mode types or create waveguides with a filtering function for bandpass.[36,37] Excitation of various types of modes is used in antenna irradiators. When connecting a waveguide to an antenna, full power transfer and the absence of signal distortion are usually required.

Can approximate methods describe the solution obtained in the article? The flexibility of the presented exactly solvable model is due to not only the possibility of choosing arbitrary material of the line $(L_0, C_0, L_1, C_1)$, but also the presence of the geometric factor, i.e. free parameters characterizing the lengths of the inhomogeneities in the non-uniform segment $(l_1, l_2, d)$. As a result, the changes cover all possible cases: the properties of the system can change both in a jump (due to the choice of different quantities $(L_0, C_0) \neq (L_1, C_1)$), and continuously; at the same time, they can change both quickly and slowly, both increase and decrease (due to the choice of quantities $(l_1, l_2, d)$). It is well known that the eikonal approach in wave physics can be applied in the cases of slow variations of parameters of wave field or media along the wavelength.[48] For the considered model, this approximation covers the very particular case only: $(L_0, C_0) = (L_1, C_1)$, $d \ll |l_1|$, $d \ll |l_2|$, $N \to 1$ (i.e. either $\tilde{\omega} \gg 1$, or $l_1 \to l_2$). Under these conditions, the eikonal approximation can allow to approximately determine the intensity of the transmitted wave. The anti-eikonal limit can be applicable for determination of the wave intensity in the opposite case, when the system parameters highly varies over the distance of one wave-length, i.e. $\lambda_0 \gg 2\pi \max\{d, l_1, l_2\}$, $\tilde{\omega} \ll 1$. This case can also include a system parameter jump: $(L_1, C_1) \to (L_0, C_0)$. However, in the general case, none of the approximations mentioned above tracks the exact changes in the wave phase. And since the implementation of reflectionless passage is determined by the exact values of the phases (namely, the phases change so that at the point of incidence all reflected waves cancel each other out in total), then none of the approximations catches resonance cases.

A non-uniform segment of the TL acts in relation to the incident wave as a wave "barrier". Each such barrier can be described by a complex transmission coefficient $T_i$ and complex reflection coefficients for a wave incident on the left $R_i$ (impinges from $-\infty$) and for a wave incident on the right $\tilde{R}_i$ (impinges from $+\infty$). Let several such non-uniform segments be included in the homogeneous TL and they are separated by the homogeneous segments of the TL with a length $D_i$. Let us start with the combined system of two such barriers. In the general case the wave amplitudes for repeatedly reflected waves between these separate barriers can be exactly summed up as the geometric progression (see Fig. 7), and the complex transmission coefficient for the combined barrier can be found:[7]

$$T_{12} = \frac{T_1 T_2}{1 - \tilde{R}_1 R_2 e^{2ik_1 D_1}}. \tag{24}$$

Similarly, summing up the return flows, we obtain the complex reflection coefficients:

$$R_{12} = R_1 + \frac{T_1^2 R_2 e^{2ik_1(d_1 + D_1)}}{1 - \tilde{R}_1 R_2 e^{2ik_1 D_1}}, \quad \tilde{R}_{21} = \tilde{R}_2 + \frac{T_2^2 \tilde{R}_1 e^{2ik_1(d_2 + D_1)}}{1 - \tilde{R}_1 R_2 e^{2ik_1 D_1}}. \tag{25}$$

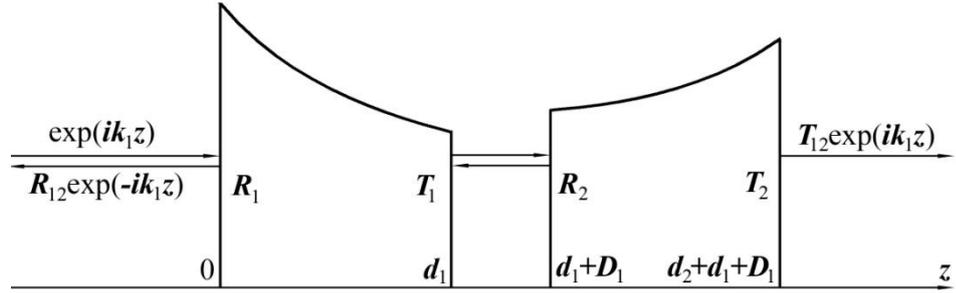

**Fig. 7.** Tunneling of a wave through a pair of barriers, each of which is characterized by its own reflection and transmission coefficients $(R_1, T_1)$ and $(R_2, T_2)$ accordingly; widths of the barriers are $d_1$ and $d_2$ respectively, $D_1$ is the distance between these barriers.

For a wave with some frequency $\omega$, the reflectionless passage ($|T_{12}| = 1$) is possible, if the following condition is met:

$$\tilde{R}_1 = R_2^* e^{-2ik_1 D_1}, \tag{26}$$

that is possible when $|R_1|^2 = |R_2|^2$ for this frequency (but the barriers themselves can be different: $R_1 \neq R_2$). In this case, if the resonant passage is realized for the first barrier $(R_1, T_1)$, then additional resonant frequencies appear when the second barrier $(R_2, T_2)$ is located at such a resonant distance $D_1$.

We consider the following example. Let the first segment of the non-uniform line with the former characteristics (6) be located between $z = d + D_1$ and $z = 2d + D_1$, and the second segment of the non-uniform line be located between $z = 0$ and $z = d$. Its characteristics, instead of expressions (6), will be the following distributed capacity $P_2(z)$ and inductance $W_2(z)$:

$$P_2(z) = \left(1 + \frac{2d + D_1 - z}{l_1}\right)^{-1}, \quad W_2(z) = \left(1 + \frac{2d + D_1 - z}{l_2}\right)^{-1}. \tag{6'}$$

Choosing in previous dimensionless designations $\tilde{l}_1 = 10$, $\tilde{l}_2 = 5$, $\tilde{Z} = 5$, the dependence of the value of $|T_{12}|$ on the dimensionless frequency $\tilde{\omega}$ can be depicted. If we take dimensionless value of $k_1 D_1 = 0.001346077$, then, as we see from Fig. 8, the number of resonant frequencies (when reflectionless transmission $|T_{12}| = 1$ is realized) will be uniformly doubled compared with the case when there was only one non-uniform segment.

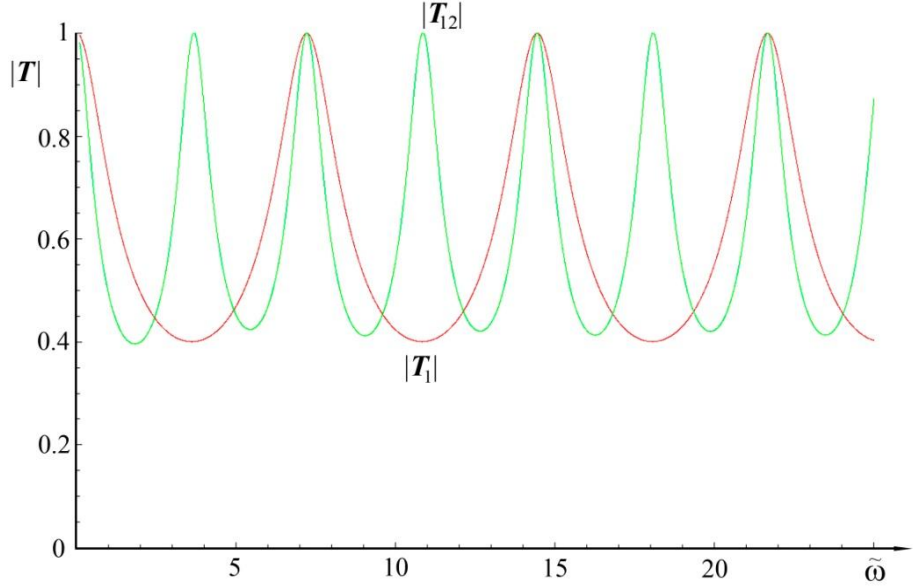

**Fig. 8.** Comparative dependence of the transmission coefficient $|T_1|$ for the single barrier at the fixed parameters $\tilde{l}_1 = 10$, $\tilde{l}_2 = 5$, $\tilde{Z} = 5$ and of the transmission coefficient $|T_{12}|$ for the pair of similar barriers (see (6) and (6')) on the dimensionless frequency $\tilde{\omega}$. The curves are computed with using the expressions (21), (24) and the resonance condition (26) for the resonance distance. The dependencies have an oscillatory character. As it can be seen from the graphs, additional resonance frequencies appeared for the transmission coefficient $|T_{12}| = 1$.

We can do otherwise: to pre-select a frequency that we want to make as the resonant one and to use such $D_1$ (in doing so, other additional resonant frequencies will appear also). For example, choosing $\tilde{\omega}_r = 4$ as the resonance frequency, we obtain the dependence of the transmission coefficient on the frequency shown in Fig. 9.

Of extra interest is the tunneling regime. Note that for an ideal TL, even if one inhomogeneous region is an opacity region for some frequency, there may be such a resonant distance that the wave can pass through two such opaque regions without reflection, so that the wave amplitude is completely restored (resonant tunneling). As we see from Fig. 10, such resonances are very sharp $(\Delta\omega \to 0)$. Here we chose the frequency $\tilde{\omega} = 1$ from opaque region ($|T_1| \to 0$) in advance and made the reflectionless passage for this frequency ($|T_{12}|=1$). For non-ideal lines, the reflectionless passage is impossible for frequencies in opaque region.

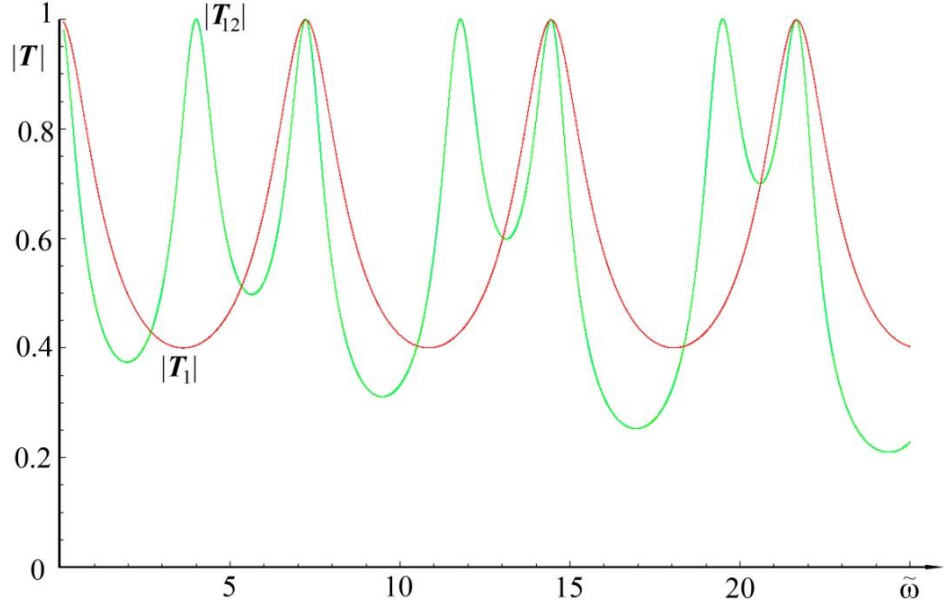

**Fig. 9.** Dependencies of the transmission coefficient $|T_1|$ for the single barrier at the fixed parameters $\tilde{l}_1 = 10$, $\tilde{l}_2 = 5$, $\tilde{Z} = 5$ and of the transmission coefficient $|T_{12}|$ for the pair of similar barriers on the dimensionless frequency $\tilde{\omega}$. Here the resonance frequency of $\tilde{\omega}_r = 4$ is taken in advance, and using the resonance condition (26) for the resonance distance, the curves are computed from the expressions (21) and (24). As it can be seen from the graphs, additional resonance frequencies appeared for the transmission coefficient $|T_{12}|=1$.

Let now the TL contains *n* non-uniform segments. Applying successively formulas (24), (25) when each new segment is attached, we can obtain the final expressions for the transmission and reflection coefficients at the *n*-th step:

$$T_{1\ldots n} = \frac{T_{1\ldots n-1} \cdot T_n}{1 - \tilde{R}_{1\ldots n-1} R_n \exp\{2ik_1 D_{n-1}\}}, \qquad (27)$$

$$R_{1\ldots n} = R_{1\ldots n-1} + \frac{T_{1\ldots n-1}^2 R_n \exp\left\{2ik_1 \sum_{j=1}^{n-1}(d_j + D_j)\right\}}{1 - \tilde{R}_{1\ldots n-1} R_n \exp\{2ik_1 D_{n-1}\}}. \qquad (28)$$

If identical non-uniform segments of the TL are arranged in pairs at such a resonant distance, then you can always arrange the pairs so that the number of resonant frequencies will increase many times.

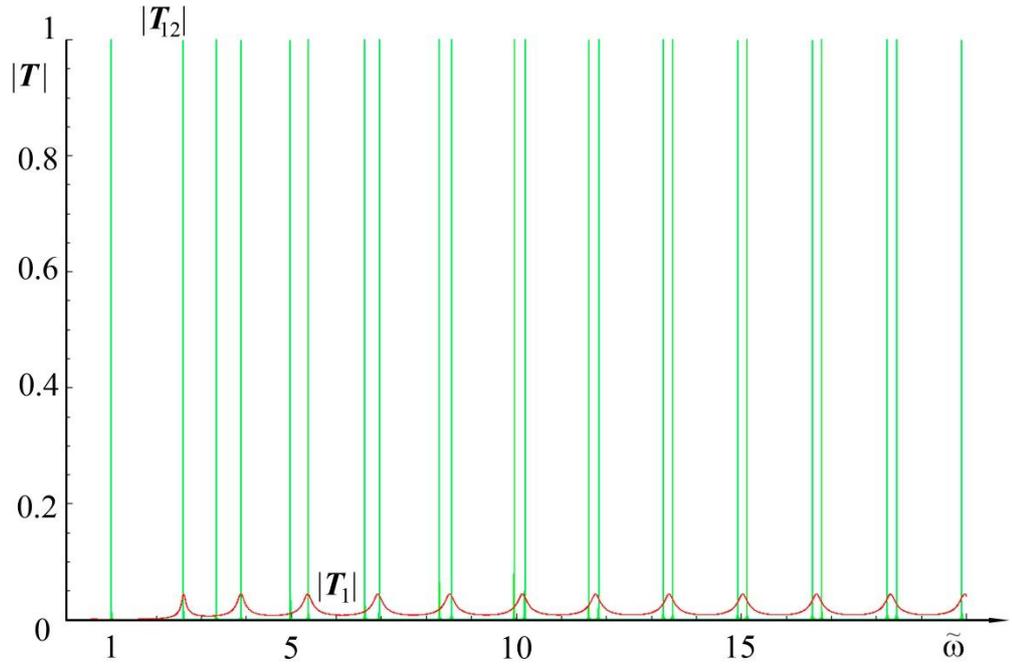

**Fig. 10.** Dependencies of the transmission coefficient $|T_1|$ for the single barrier at the fixed parameters $\tilde{l}_1 = 1$, $\tilde{l}_2 = -0.999$, $\tilde{Z} = 5$ and of the transmission coefficient $|T_{12}|$ for the pair of similar barriers on the frequency $\tilde{\omega}$. Here the resonance frequency of $\tilde{\omega}_r = 1$ is chosen in advance from the opaque region. Using the resonance condition (26) for the resonance distance, these curves are computed from the expressions (21) and (24). As it can be seen from the graphs, previously missing resonance frequencies appear for the transmission coefficient $|T_{12}| = 1$, including in the opacity region.

## IV. CONCLUSION

The present work was devoted to the problems of the appearance of wave dispersion due to the TL inhomogeneity. The exactly solvable model discussed in this paper helps us to better understand the physical process of signal passing through a non-uniform section of the line and to compare the exact solutions derived here and those obtained using various approximations. The proposed approach based on the use of new variables (new phase coordinate $\eta$ and new function) made it possible to construct exact analytical solutions to telegraph equations with a continuous distribution of parameters depending on the coordinate. The flexibility of the discussed model is due to the presence of a number of free parameters, including two geometric parameters characterizing the lengths of the inhomogeneities in values of *L* and C. In the new variables, the spatiotemporal structure of the solutions is described using sine waves and elementary functions, and the dispersion is determined by the formulas of the waveguide type. The rigorous expressions for the complex reflection and transmission coefficients were derived. These expressions describe not only amplitudes, but also phase shifts in reflected and transmitted waves. The following interesting cases were analyzed: the passage of waves without phase change, the reflectionless passage of waves, and the passage of signals through a sequence of non-uniform sections.

Continuous modulation of parameters of the line allows us to provide following new properties for a long TL:

- waveguide-type frequency dispersion, which is characterized by a controlled cutoff frequency;
- the possibility of coordination (adjustment) for sections of the line, regardless of their geometric characteristics;
- transmitted wave phase control.

An important feature of such lines is the fact that the refractive index can be arbitrary (more than one, less than one, and even imaginary). For a given spectral range, the line parameters can be set in such a way that non-local dispersion effects will appear in the required frequency band. In this case, the wave dynamics in the line is described by exact analytical solutions constructed without any assumptions about the smallness of the changes in the parameters of the line or fields.